\documentclass[reprint,groupedaddress,amsmath,amssymb,aps,pra,floatfix]{revtex4-1}

\pdfoutput=1
\usepackage{amsmath,amssymb,graphicx,amsfonts,braket,comment,enumitem}

\newcommand{\bs}[1]{\boldsymbol{#1}}

\begin{document}

\title{Sawtooth Wave Adiabatic Passage in a Magneto-Optical Trap}

\author{John P. Bartolotta}
\affiliation{JILA, NIST, and University of Colorado, 440 UCB, 
Boulder, CO  80309, USA}
\author{Murray J. Holland}
\affiliation{JILA, NIST, and University of Colorado, 440 UCB, 
Boulder, CO  80309, USA}
\email[]{john.bartolotta@colorado.edu}

\date{\today}

\begin{abstract}
\noindent We investigate theoretically the application of Sawtooth Wave Adiabatic Passage (SWAP) in a 1D magneto-optical trap (MOT). As opposed to related methods that have been previously discussed, our approach utilizes repeated cycles of stimulated absorption and emission processes to achieve both trapping and cooling, thereby reducing the adverse effects that arise from photon scattering. Specifically, we demonstrate this method's ability to cool, slow, and trap particles with fewer spontaneously emitted photons, higher forces and in less time when compared to a traditional MOT scheme that utilizes the same narrow linewidth transition. We calculate the phase space compression that is achievable and characterize the resulting system equilibrium cloud size and temperature.
\end{abstract}

\pacs{}

\maketitle 


\section{INTRODUCTION} \label{intro}
\noindent Trapping and slowing devices based on the removal of momentum, energy, and entropy via light are ubiquitous and essential in most experiments on quantum gases. Techniques such as the slowing and cooling of particles by preferential absorption from a counterpropagating laser~\cite{firstRadPressure,dopp1}, the magneto-optical trap~\cite{mot}, Zeeman decelerators~\cite{zeemanSlower}, the bichromatic force \cite{bichromatic}, and sawtooth-wave adiabatic passage (SWAP) cooling~\cite{exp,swap_theory} all rely on many cycles of an engineered light-matter interaction for this purpose. 

It is widely understood that spontaneous emission is a fundamental requirement for the laser cooling and trapping of atoms and molecules. This is due to the fact that the scattered photons irreversibly remove entropy, allowing the system to violate the conditions for Liouville's theorem to apply and to thereby undergo compression in terms of the occupied volume in phase space. While incorporating repeated scattering events is acceptable for systems with a closed cycling transition, this can lead to significant loss for systems with many degrees of freedom, such as molecules, which may have a large number of dynamically decoupled internal states, or ``dark states." Protocols such as the optical repumping from uncoupled states to cooled states can mitigate this issue~\cite{Kastler,shuman_barry_demille_2010}, but a more absolute solution would be to significantly reduce the number of spontaneous emissions necessary to achieve slowing, cooling, and trapping. 

SWAP cooling can cool a system to equilibrium with fewer spontaneous emissions than Doppler cooling and slow a distribution of particles using purely coherent dynamics~\cite{swap_theory}. We present the results of combining SWAP cooling with a quadrupole magnetic trap, which we call a SWAP MOT~\cite{norciaThesis}. We show that a SWAP MOT is able to demonstrate slowing, cooling, and trapping with fewer scattered photons, higher conservative forces, and in less time when compared with a traditional MOT scheme.  

\begin{figure}
    \centering
    \includegraphics[width=0.85\linewidth]{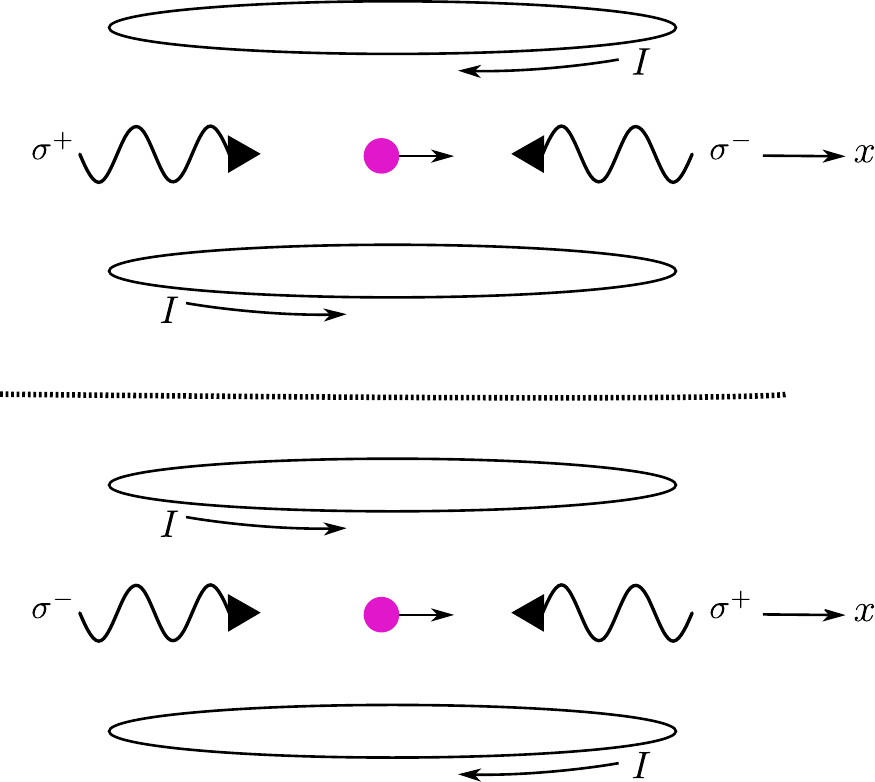}
    \caption{A schematic of the two experimental setups for the 1D SWAP MOT (motion only along the $x$-direction) that operate in alternating periods of the cooling cycle. Anti-Helmholtz coils with current $I$ and counterpropagating lasers of opposite circular polarization $\sigma^+$ and $\sigma^-$ create a magnetic-optical trap for neutral particles (pink circle). The laser polarizations and current directions are (ideally) instantaneously, periodically switched between the two setups as the detuning of each laser, which is linearly ramped from below to above the resonance of the cooling transition, passes through zero.}
    \label{fig:expSetup}
\end{figure}

Our approach differs from other methods of implementing the SWAP procedure in a MOT~\cite{snigirev, silva} because we formulate a method to incorporate the stimulated emission process originally envisioned in the first proposals of SWAP cooling~\cite{exp,swap_theory}. This addition promotes both cooling and trapping and reduces the number of scattered photons required for equilibration. In order to achieve the desired coherent dynamics, we propose that the cooling laser polarizations and magnetic field directions (or, equivalently, the direction of the electron's magnetic moment) should be abruptly switched at the center and end of each sweep (see Fig.~\ref{fig:expSetup}).

In Section \ref{mechanism} we explicate the particle dynamics achieved with the SWAP MOT protocol. In Section \ref{model} we develop a semiclassical model in which the internal states are treated quantum mechanically and the external states are treated classically. Section \ref{quantum_traj} explains the details of our numerical algorithm. In Section \ref{phasespace} we define regimes in phase space that exhibit different types of dynamic behavior under SWAP MOT evolution, define capture range conditions, and demonstrate phase space compression and MOT loading. In Section \ref{equilibrium} we provide various scaling properties of the procedure over the range of interesting system parameters.


\section{SWAP MOT mechanism} \label{mechanism}

\begin{figure}
    \centering
    \includegraphics[width=0.7\linewidth]{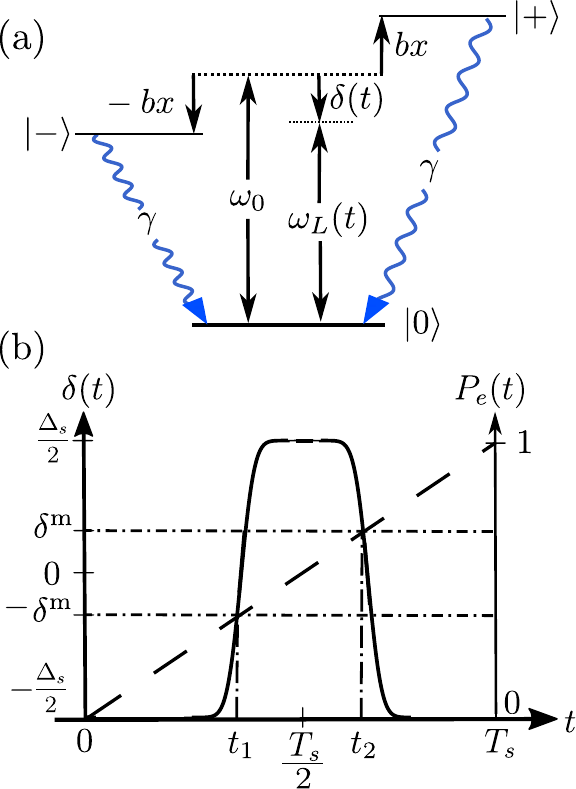}
    \caption{(a) The minimum internal state structure necessary for demonstrating SWAP MOT dynamics. The transition frequency $\omega_0$, natural linewidth $\gamma$, laser frequency $\omega_L(t)$, detuning $\delta(t)$, and Zeeman shifts $\pm bx$ of the excited states are included. (b) The laser detuning $\delta(t)$ (dashed) and excited state fraction $P_e(t)$ (solid) vs time $t$ over a sweep of period $T_s$.
    The particle resonates with the lasers at the times $t_1$ and $t_2$ when $\delta(t) = \pm \delta^\text{m}$ (dot-dashed), where $\delta^\text{m}$ is the motional detuning [see Eq.~(\ref{eq:motionaldetuning})].}
    \label{fig:mechanism}
\end{figure}
The SWAP procedure relies on the coherent transfer of a particle between quantum states via adiabatic rapid passage.
The internal structure of the simplest quantum system that demonstrates the desired dynamics consists of two excited states with a common ground state, which we label $\ket{+}, \ket{-},$ and $\ket{0}$, respectively.
The labels correspond to the values $m=\pm1$ and $0$, where $m$ is the magnetic quantum number, such as that occurs in a system with a  $J=0\rightarrow J=1$ transition, where $J$ is an angular momentum quantum number. Due to our choice of laser polarization, omission of the $\ket{J,m} = \ket{1,0}$ state is valid, as it is not optically pumped.
As shown in Fig.~\ref{fig:mechanism}(a), the excited states are separated in energy from the ground state by $\hbar \omega_0$, where $\omega_0$ is the transition frequency, assumed to be in the optical domain. In the presence of a magnetic field, the excited states are shifted in energy according to the Zeeman shift $\pm \hbar b x$, where $b= g \mu_\text{B} (\nabla B)/\hbar$, $g$ is the g-factor of the transition, $\mu_\text{B}$ is the Bohr magneton, and $x$ is the displacement of the particle from the trap center. 
Both excited states decay to the ground state with a rate given by the linewidth $\gamma$. 
For simplicity, we limit our discussion to one dimension.

As displayed in Fig.~\ref{fig:expSetup}, the experimental setup is nearly identical to that of a type I 1D MOT \cite{metcalf}. However, instead of fixing the laser detunings
\begin{equation}
\label{eq:laserDetuning}
    \delta(t)=\omega_L(t) -\omega_0
\end{equation}
below the transition frequency, they are repeatedly swept from below to above the cooling transition in a sawtooth pattern (see the dashed curve in Fig.~\ref{fig:mechanism}(b)) with full period $T_s$. In Eq.~\eqref{eq:laserDetuning}, $\omega_L(t)$ is the instantaneous laser frequency. Additionally, the directions of the currents $I$ in the anti-Helmholtz coils and the polarizations of the cooling lasers are switched at a rate $2/T_s$.  The current directions are chosen such that the magnetic field along the radial direction at the center of the trap, which we call the $x$-direction (see Fig.~\ref{fig:expSetup}), has the form $B(x) = (\nabla B) x$ during the first half of the sweep, and $B(x) = -(\nabla B) x$ during the second half, where $\nabla B >0$ is the magnetic field gradient. The laser traveling along the $+x$ ($-x$)-direction has circular polarization $\sigma^+$ ($\sigma^-$)  during the first half, and then these polarizations are exchanged for the second half.

The general desired coherent dynamics over a single cycle of the SWAP procedure is as follows. Let us assume that the particle begins in $\ket{0}$, which is a good assumption due to the effects of spontaneous emission for appropriate system parameters. The particle resonates with one of the lasers when the laser detuning $\delta(t)$ is equal in magnitude to the particle's motional detuning
\begin{equation}
\label{eq:motionaldetuning}
\delta^\text{m} \equiv bx + kv,
\end{equation}
which is the sum of its Doppler and Zeeman shifts, as shown in Fig.~\ref{fig:mechanism}(b). Here, $k$ is the wavenumber of the transition and $v$ is the velocity of the particle. More specifically,  at the time $t_1$ during the first half of the sweep defined by $\delta(t_1) = - |\delta^\text{m}|$, the particle absorbs a photon from one of the cooling lasers and is transferred into whichever excited state $\ket{e} \in \{ \ket{+}, \ket{-} \}$ first comes into resonance. Then, at the time $t_2$ in the second half of the sweep defined by $\delta(t_2) = |\delta^\text{m}|$, the particle emits a photon into the other laser by stimulated emission and is transferred coherently back to $\ket{0}$ [see $P_e(t)$ in Fig.~\ref{fig:mechanism}(b)] with the corresponding momentum shift. It is essential to our method that both the laser polarizations and magnetic field direction are switched between the times $t_1$ and $t_2$ so that the particle resonates with the correct laser as to achieve stimulated emission and the consequent second momentum recoil.
This second stimulated process in a sweep is exactly what differentiates our method from other SWAP MOT protocols~\cite{snigirev, silva}.
Very importantly, this protocol replaces the scattering event required after every absorption in Doppler cooling with a stimulated emission, mitigating the adverse effects of momentum diffusion that would otherwise occur. This feature allows the particle to experience the impulse of many photon momenta while avoiding spontaneous emission.

\begin{figure*}
    \centering
    \includegraphics[width=0.9\linewidth]{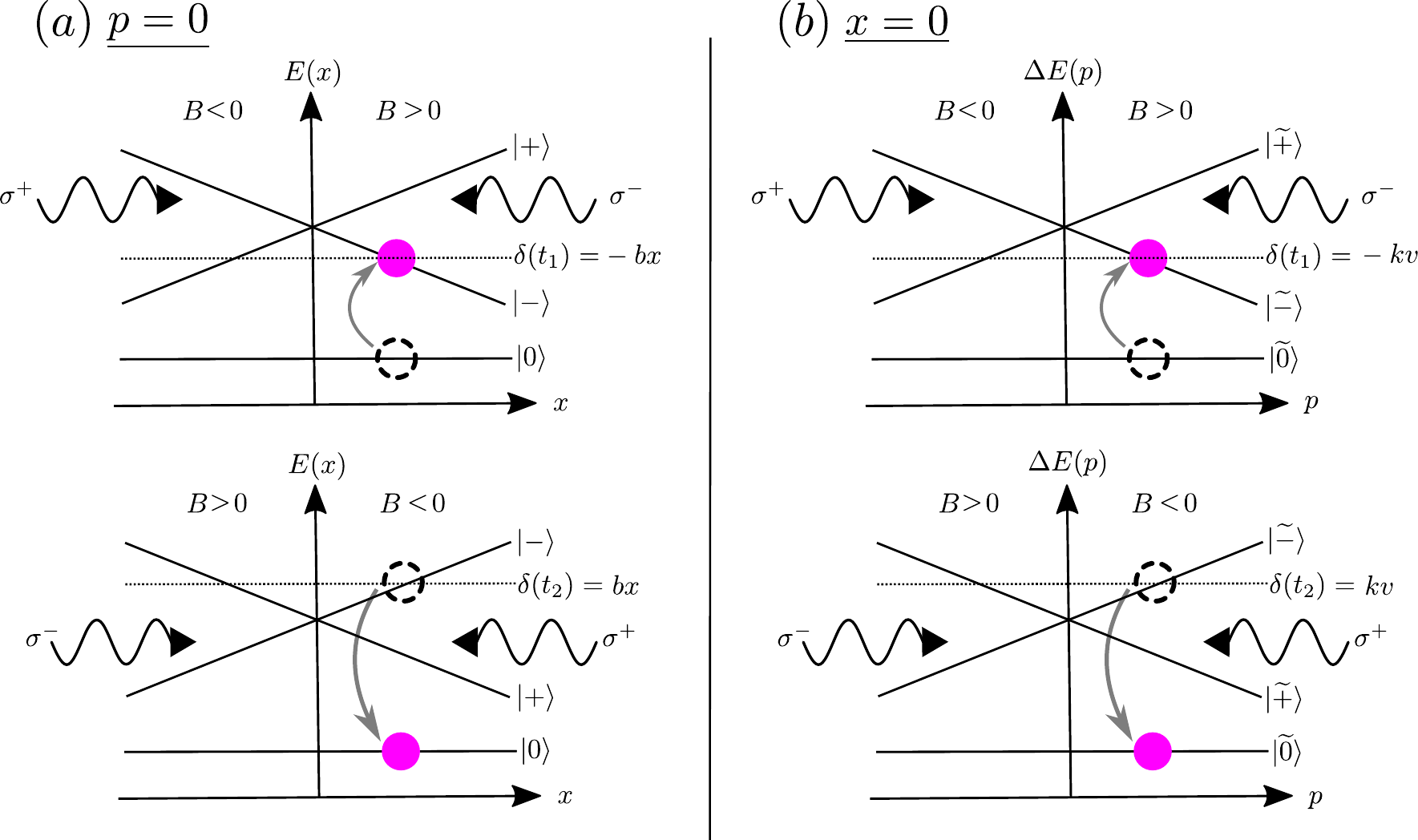}
    \caption{ A qualitative demonstration of the trapping force (a) and slowing force (b) in a SWAP MOT. The cooling lasers (with circular polarizations $\sigma^\pm$) and sign of the magnetic field $B(x)$ setup are included. (a) Top: The energy eigenvalues $E(x)$ of $\hat{H}(t)$ with $\Omega = 0$ of a motionless particle ($p=0$) as a function of position $x$ during the first half of the sweep. An example system (pink circle) with $x >0$ in the state $\ket{0}$ absorbs a $\sigma^-$ photon, yielding an impulse toward $x=0$ and transferring the particle into the $\ket{-}$ state. This occurs at the time $t_1$ when the swept laser frequency satisfies $\delta(t_1)=-bx$. Bottom: $E(x)$ vs. $x$ and laser and magnetic field setup during the second half of the sweep. The example system emits a $\sigma^-$ photon, yielding another impulse toward $x=0$ and transferring the particle back to the $\ket{0}$ state. This occurs at the time $t_2>t_1$ when the swept laser frequency satisfies $\delta(t_2)=bx$. (b) Top: The relative energy eigenvalues $\Delta E(p) \equiv E(p)- \tfrac{p^2}{2m}$ of of $\hat{H}(t)$ with $\Omega = 0$ for the states $W(p)$ [see Eqn.~(\ref{eq:subset})] for a particle at the center of the trap ($x=0$) as a function of momentum $p$ in the first half (top) and second half (bottom) of the sweep. An example system experiences an impulse toward $p=0$ by undergoing similar dynamics to Fig.~\ref{fig:energy}(a) with the substitutions $x\rightarrow p$ and $bx \rightarrow kv$.}
    \label{fig:energy}
\end{figure*}

It is necessary to demonstrate that this protocol inherently generates a force toward the center of the trap (trapping), a force that opposes the particle's motion (slowing),  and an overall frictional force (cooling). In order to illustrate its trapping capability, consider a motionless particle with position $x>0$ [see Fig.~\ref{fig:energy}(a)]. The Zeeman shift causes the particle to absorb a $\sigma^-$ photon from the left-traveling laser during the first half of the cycle and to emit a photon into the right-traveling $\sigma^-$ laser during the second half, transferring the particle back to $\ket{0}$ with a net impulse of two photon momenta toward the center of the trap. Its slowing capability is elucidated by considering a particle with momentum $p>0$ near the center of the trap [see Fig.~\ref{fig:energy}(b)]. In this case, the Doppler shift causes the particle to absorb a $\sigma^-$ photon from the left-traveling laser during the first half and to emit a photon into the right-traveling $\sigma^-$ laser during the second half, transferring the particle back to $\ket{0}$ also with a net impulse of two photon momenta that opposes the particle's motion. Examples with negative $x$ or $p$ similarly would cause trapping and slowing via interaction with the $\sigma^+$ laser. The cooling effect is more subtle. To be precise, we define {\em cooling}\/ as compression of the particle's classical phase space volume, which is achieved through irreversible entropy flow from the particle to free space via the process of spontaneous emission. This most frequently occurs when the Doppler and Zeeman shifts are equal in magnitude but opposite in sign, as the particle resonates with both lasers simultaneously and is left with a significant excited state fraction at the end of a sweep. Spontaneous emission then resets the particle to the ground state for the next sweep, yielding a net drift in phase space toward the phase space origin. Although the discussion so far has been descriptive, we we will elucidate the details of the full dynamics in Section \ref{phasespace} through the numerical solutions to follow.

\section{Model and Semiclassical Approximations} \label{model}

\noindent In order to capture the intricate features of the SWAP procedure, we first develop a fully quantum mechanical model and then make appropriate semiclassical approximations to create a computationally tractable simulation. As previously described, Fig.~\ref{fig:mechanism}(a) displays the internal state structure. We track motion along one dimension, for which the particle has momentum and position operators $\hat{p}$ and $\hat{x}$. 

The system is subject to counter-propagating lasers of opposite circular polarization (which we denote as $\sigma^+$ and $\sigma^-$) of instantaneous frequency $\omega_L(t)$ and a magnetic field that depends linearly on the coordinate as $\hat{B}(\hat{x})= (\nabla B) \hat{x}$. The detunings of the lasers $\delta(t)$ [see Eq.~\eqref{eq:laserDetuning}] are set to follow a sawtooth waveform pattern centered at zero with range $\Delta_s$ and period $T_s$. In the Schr\"{o}dinger picture, the coherent dynamics is described by the Hamiltonian
\begin{align}
    \hat{H}(t)  = & \frac{\hat{p}^2}{2m} 
                    + \hbar \omega_0 \left( \ket{+} \bra{+}
                    + \ket{-}\bra{-} \right) \notag \\
               & + \hbar b \hat{x} \left(\ket{+}\bra{+} 
                    - \ket{-}\bra{-} \right) \notag \\
               & + \frac{\hbar \Omega}{2} \left( e^{i[k\hat{x} - \eta(t)]} \hat{\sigma}^+_+ + h.c.        \right) \notag \\
               & + \frac{\hbar \Omega}{2} \left( e^{-i[k\hat{x} + \eta(t)]} \hat{\sigma}^+_- + h.c.        \right)
\end{align}

\noindent during the first half of the sweep, with similar form but substitutions $k \rightarrow -k$ and $b \rightarrow -b$ during the second half of the sweep. The particle mass is $m$, the wavenumber of the laser light (which we approximate to be constant) is $k$, and the magnetic field gradient is characterized by $b$, which was defined in Section~\ref{mechanism}. The (equal) Rabi frequencies of both of the counterpropagating lasers are~$\Omega$, and 
\begin{equation}
\eta(t) \equiv \int_0^t \omega_L(t') \, dt'
\end{equation}

\noindent is the time-dependent accumulated phase of the laser field. The operators $\hat{\sigma}^+_i \equiv \ket{i}\bra{0}$ and $\hat{\sigma}^-_i \equiv (\hat{\sigma}^+_i)^\dag = \ket{0}\bra{i}$ are the raising and lowing operators corresponding to transitions between the ground state and the two excited states. 

The quantum master equation
\begin{equation}
\label{eq:master}
 \frac{d \hat{\rho}}{dt} = \frac{1}{i \hbar} \left[ \hat{H}, \hat{\rho} \right] + \hat{\mathcal{L}}(\hat{\rho})
\end{equation}

\noindent governs the time evolution of the density matrix, $\hat{\rho}$, which fully describes the particle's internal and external states. The Lindblad superoperator, 
\begin{align}
\label{eq:lindblad}
  \hat{\mathcal{L}}(\hat{\rho}) & =
  - \frac{\gamma}{2} \sum_{i \in \left\{ +, - \right\}}
    \Bigl[
    \hat{\sigma}^+_i \hat{\sigma}^-_i \hat{\rho}
    + \hat{\rho} \hat{\sigma}^+_i \hat{\sigma}^-_i
    - 2
    	\Bigl( 
        \frac{3}{5} \hat{\sigma}^-_i \hat{\rho} \hat{\sigma}^+_i \notag \\
    & + \frac{1}{5}e^{i k \hat{x}} \hat{\sigma}^-_i \hat{\rho} \hat{\sigma}^+_i e^{- i k \hat{x}}
    	+ \frac{1}{5}e^{- i k \hat{x}} \hat{\sigma}^-_i \hat{\rho} \hat{\sigma}^+_i e^{ i k \hat{x}}
    	\Bigr)
    \Bigr],
\end{align}

\noindent captures the incoherent dynamics due to spontaneous emission and its associated recoil. We have approximated the continuous dipole radiation pattern to produce discrete recoils of magnitudes $-\hbar k,0$, and $\hbar k$ along the $x$-direction with probabilities of $\tfrac{1}{5}:\tfrac{3}{5}:\tfrac{1}{5}$, respectively~\cite{castin,dumzoller}. Although it is not absolutely necessary to make the simplification that there are only three allowed impulses arising from the direction of the photon emission, the resulting numerical implementation is more straightforward due to the fact that one may take advantage of the presence of uncoupled momentum families.

In the interaction picture defined by the free Hamiltonian
\begin{equation}
    \hat{H}_0(t) = \hbar \omega_L(t) \left(\ket{+} \bra{+}
                    + \ket{-}\bra{-} \right),
\end{equation}

\noindent the interaction Hamiltonian is
\begin{align}
\label{eq:interactionHam}
    \hat{H}_\text{I}(t)  = & \frac{\hat{p}^2}{2m} 
                    - \hbar \delta(t) \left( \ket{+} \bra{+} + \ket{-}\bra{-} \right) \notag \\
               & + \hbar b \hat{x} \left(\ket{+}\bra{+}- \ket{-}\bra{-} \right) \notag \\
               & + \frac{\hbar \Omega}{2} \left( e^{i k \hat{x}} \hat{\sigma}^+_+ + h.c.\right) \notag \\
               & + \frac{\hbar \Omega}{2} \left( e^{-i k \hat{x}} \hat{\sigma}^+_- + h.c. \right).
\end{align}

\noindent The corresponding force operator that describes the coherent evolution of the external variables is then
\begin{align}
\label{eq:quantumForce}
    \hat{F}  \equiv & - \frac{\partial \hat{H}_\text{I}}{\partial \hat{x}} \notag \\
        = \, & \hbar b \left(\ket{-}\bra{-}- \ket{+}\bra{+} \right) \notag \\
        & - \frac{i \hbar \Omega k }{2} \left( e^{i k \hat{x}} \hat{\sigma}^+_+ - h.c.\right) \notag \\
        & + \frac{i \hbar \Omega k}{2} \left( e^{-i k \hat{x}} \hat{\sigma}^+_- - h.c. \right).
\end{align}

We now make several approximations to form a tractable, semiclassical model. The parameter that characterizes the separation of timescales required to perform a typical semiclassical approximation is \cite{Dalibard_1985}
\begin{equation}
\label{eq:chi}
    \chi \equiv \frac{\omega_r}{\gamma} \ll 1,
\end{equation}
where $\omega_r \equiv \hbar k^2/2m$ is the recoil frequency of the transition. However, the SWAP procedure causes fewer spontaneous emissions as $\chi$ increases \cite{swap_theory}, so Eq.~(\ref{eq:chi}) may not necessarily hold. Moreover, the rapid adiabatic passage protocol generates coherences between quantum states, so tracking a single value for the semiclassical momentum $p$ is insufficient. Instead, we simplify our formalism by limiting the number of quantum states we track at any given time. We retain the term ``semiclassical" for this simplification and our consequent treatment of the particle's external variables.

The $\sigma^+ - \sigma^-$ polarization scheme prevents multiphoton processes by angular momentum conservation \cite{Dalibard_1985}, so a particle in an eigenstate $\ket{\widetilde{0}(p)} \equiv \ket{0,p}$ can only be transferred to the states $\ket{\widetilde{-}(p)} \equiv \ket{-,p-\hbar k}$ and $\ket{\widetilde{+}(p)} \equiv \ket{+,p+\hbar k}$ (and vice versa) during the first half of the sweep. The same statement holds true for the second half of the sweep with $k\rightarrow -k$. Therefore, we only track the particle's evolution in the subset of basis states 
\begin{equation}
\label{eq:subset}
  W(p) \equiv 
  \{
  \ket{\widetilde{0}},
  \ket{\widetilde{+}},
  \ket{\widetilde{-}}
  \}
\end{equation}
of the composite Hilbert space, then update $p$ appropriately as the particle absorbs and emits photons. 

Next, we treat the particle's external variables in the Hamiltonian $\hat{H}_I(t)$ semiclassically by making the substitutions $\hat{p} \rightarrow p$ and $\hat{x} \rightarrow x$, where $p$ is a momentum eigenvalue satisfying $\hat{p} \ket{p} = p \ket{p}$ [which also parameterizes $W(p)$], and  $x$ is a c-number that characterizes the particle's spatial position. These classical phase space coordinates will be updated according to a procedure that we develop in Section~\ref{quantum_traj}. With the state $\ket{\widetilde{0}}$ defining zero energy, the semiclassical Hamiltonian that evolves $W(p)$ is
\begin{align}
\label{eq:SCH}
    \hat{H}_\text{s}(t) = & -\hbar (\delta(t) - \delta^\text{m}) \ket{\widetilde{+}} \bra{\widetilde{+}} \notag \\
                          &  -\hbar (\delta(t) + \delta^\text{m}) \ket{\widetilde{-}}\bra{\widetilde{-}} \notag \\
               & + \frac{\hbar \Omega}{2} \left( \hat{\sigma}^x_{_{\widetilde{+}}} + \hat{\sigma}^x_{_{\widetilde{-}}} \right),
\end{align}

\noindent where $\hat{\sigma}^x_{_{\widetilde{i}}} \equiv \hat{\sigma}^+_{_{\widetilde{i}}} + \hat{\sigma}^-_{_{\widetilde{i}}}$. Note that the motional detuning $\delta^\text{m}$ determines the condition for resonance in Eq.~(\ref{eq:SCH}) as previously discussed. Additionally, we have omitted recoil energy terms $\hbar \omega_r$ under the approximation that they are small in comparison to $\hbar \delta^\text{m}$. 

The coherent evolution of $x$ and $p$ is typically accounted for via Newton's laws of motion under the classical force
\begin{align}
\label{eq:classicalForce}
    F(t) &= \left\langle \hat{F}(t) \right\rangle \notag \\
    & = \hbar b \left[
    \left| c_{_{\widetilde{-}}}(t) \right|^2
    -\left| c_{_{\widetilde{+}}}(t) \right|^2 \right] \notag \\
      & \quad  +  \hbar \Omega k \left( 
           \text{Re}\left[ i \langle \hat{\sigma}_{_{\widetilde{-}}}^+(t) \rangle \right] 
           -\text{Re}\left[ i \langle \hat{\sigma}_{_{\widetilde{+}}}^+(t) \rangle \right]   \right), 
\end{align}
where $\left|c_{_{\widetilde{i}}}(t)\right|^2 \equiv \left| \braket{\widetilde{i}|\Psi(t)} \right|^2$ are the instantaneous populations of state~$\ket{\widetilde{i}}$ determined from the particle's wave function $\ket{\Psi(t)}$ evolved under $\hat{H}_\text{s}(t)$. As opposed to other semiclassical treatments of laser cooling \cite{metcalf}, the effects of $F(t)$ on the particle momentum are intrinsically incorporated into our model due to the $p$-dependence of the quantum states in $W(p)$. Nevertheless, we provide this force analysis as a consistency check.
As we will motivate in Section~\ref{phasespace}, we make the physically reasonable assumption that the coherent force due to photon absorption and emission is much larger than the coherent Zeeman force that arises from the gradient of the magnetic field, {\em i.e.}, $b \ll \Omega k$, which reduces Eq.~(\ref{eq:classicalForce}) to
\begin{equation}
\label{eq:classicalForceNoB}
    F(t) \approx  \hbar \Omega k \left( 
          \text{Re}\left[ i \langle \hat{\sigma}_{_{\widetilde{-}}}^+(t) \rangle \right] 
           -\text{Re}\left[ i \langle \hat{\sigma}_{_{\widetilde{+}}}^+(t) \rangle \right]
           \right).
\end{equation}
In this limit, the impulse experienced by the particle at any time in the first half of the sweep is
\begin{align}
    \Delta p(t) & = \int_{t_0}^t F(t') \, dt' \notag \\
    \label{eq:deltaP}
    & \approx \hbar k \left[ \Delta P_+(t_0,t) - \Delta P_-(t_0,t) \right], 
\end{align}
where $\Delta P_\pm(t_0,t) \equiv P_\pm(t)-P_\pm(t_0)$ are the changes in population of the states $\ket{\widetilde{\pm}}$ between the times $t_0$ and $t$. Eq.~(\ref{eq:deltaP}) agrees with the result obtained from calculating
\begin{equation}
    \Delta p(t) = \braket{\Psi(t)|\hat{p}|\Psi(t)}-\braket{\Psi(t_0)|\hat{p}|\Psi(t_0)}.
\end{equation}

Eqns.~(\ref{eq:SCH}) and (\ref{eq:deltaP}) uniquely determine the coherent evolution of a particle during the first half of a sweep. The corresponding equations for the second half of the sweep are found by the substitutions $k \rightarrow -k$ and $b \rightarrow -b$. The incoherent dynamics are described by the Lindblad superoperator $\mathcal{L}(\hat{\rho})$ in Eq.~(\ref{eq:lindblad}) with the appropriate substitutions. 

\section{Variant of quantum trajectories}
\label{quantum_traj}

\noindent We are now in a position to choose a simulation method for determining the semiclassical evolution of particles in the SWAP MOT. In this section, we develop a variant of the Monte Carlo quantum trajectories method in the time domain~\cite{castin,dumzoller}. This choice is motivated by this method's ability to create a time record for the emission of spontaneous photons by individual particles, i.e., a quantum jump. We will use this knowledge about the jump time to simulate the motional dynamics as accurately and efficiently as possible within our semiclassical framework.

In the quantum trajectory method, the density matrix $\hat{\rho}(t)$ is obtained by averaging the associated pure states of many ``quantum trajectories," each described by a state vector~$\ket{\psi}$:
\begin{equation}
    \hat \rho(t) \approx \frac{1}{N} \sum_{i=1}^N 
    \ket{\psi(t)}\bra{\psi(t)},
\end{equation}
where $N\gg1$ is the number of simulated trajectories. The state vectors are continuously evolved under a non-Hermitian Hamiltonian defined as
\begin{equation}
    \hat{H}_\text{eff} \equiv 
    \hat{H}
    - \frac{i \hbar}{2}
    \sum_i
    \hat{J_i}^\dag
    \hat{J_i}
\end{equation}
in addition to discrete quantum jumps that correspond to the various decay channels in the system \cite{castin,dumzoller}. As we present below, our method retains these features and introduces fictitious quantum jumps of $\ket{\psi}$ into one of the elements of $W(p)$, directly followed by an update of the coordinates $x$ and $p$ based on the results of the projection. These additional operations allow us to calculate an effective, coarse-grained, single phase space trajectory $\bs{(}x(t),p(t)\bs{)}$ for each quantum trajectory. 

In the case of distinguishable spontaneous emission processes, the jump operators $\hat{J_i}$ take the form $\hat{J_i}=\sqrt{\gamma} \hat{\sigma}^-_i$. In our model, the spontaneous photons due to the decay of the the excited states $\ket{\pm}$ to the ground state $\ket{0}$ are distinguishable in principle by their polarization, so our effective semiclassical Hamiltonian is
\begin{equation}
\label{eq:hEff}
    \hat{H}_\text{eff}(t) \equiv 
    \hat{H}_\text{s}(t)
    - \frac{i \hbar \gamma}{2}
    \left(
     \ket{\widetilde{+}} \bra{\widetilde{+}} + \ket{\widetilde{-}}\bra{\widetilde{-}}
    \right).
\end{equation}

The time evolution protocol for each quantum trajectory is as follows. Note that $k>0$ ($k<0$) in the first (second) half of the sweep.

\begin{enumerate}
    \item Initialize the classical phase space coordinate $(x,p)$ by sampling from a chosen distribution. Using this choice of $p$, prepare the quantum state in a state vector that lies in the basis $W(p)$. Record the quantum state as $\ket{\psi_r}$, the time as $t_r$, and the initial phase space coordinate as $(x_r,p_r)$.
    \item \label{evolve} Continuously evolve $\ket{\psi(t)}$  under $\hat{H}_\text{eff}(t)$ [Eq.~(\ref{eq:hEff})] with the external coordinate, $(x_r,p_r)$, held constant. This evolution can be interrupted at at a time $t_j$ when one of two possible discrete quantum jump events occur:
    \begin{description}
    \item[Case A] The middle or end of a sweep is reached.
    \begin{enumerate}[label=(\roman*)]
    \item Simulate a fictitious quantum jump of $\ket{\psi(t_j)}$ into one of the basis states  in the subspace $W(p_r)$ using populations to determine the relative likelihood, i.e., $\ket{\psi(t)} \rightarrow \ket{\psi_P} \in W(p_r)$. 
    \item{\label{PSupdate} Calculate the corresponding impulse $\Delta p$ and displacement $\Delta x$ according to
    \begin{align}
    \hspace{50pt minus 1fil} 
      \Delta p & =\braket{\psi_P|\hat{p}|\psi_P}-\braket{\psi_r|\hat{p}|\psi_r}  \hfilneg \notag \\
      & \in \{0, \hbar k, - \hbar k \}; \notag \\
      \Delta x  &= \frac{1}{m} \int p(t') \, dt' 
         \approx \frac{1}{m} \frac{p_\text{avg}}{2}\Delta t \notag \\
        &=\frac{1}{m}\left(p_r + \frac{\Delta p}{2}\right)(t_j-t_r)
        \label{eqns:deltas}
    \end{align}
    so that the new phase space coordinate is
    \begin{equation}
    \hspace{50pt minus 1fil} 
    (x,p) = (x_r + \Delta x, p_r + \Delta p).\hfilneg 
    \end{equation}}
    \item Relabel the basis subset $W(p_r)$ with the updated momentum $p$. 
    \item Replace $\ket{\psi_r},t_r$ and $(x_r,p_r)$ with their updated values $\ket{\psi_P}$, $t_j$, and $(x,p)$.
    \end{enumerate}
    \item[Case B] A spontaneous photon is detected according to the quantum trajectory procedure~\cite{castin,dumzoller}.
    \begin{enumerate}[label=(\roman*)]
    \item Simulate which state ${\ket{\widetilde{e}}\in\{\ket{\widetilde{+}},\ket{\widetilde{-}}\}}$ the system occupied and hence decayed from using probabilities weighted by the excited state populations, i.e., project ${\ket{\psi(t_j)}\rightarrow\ket{\psi_P}=\ket{\widetilde{e}}}$.
    \item Apply $\Delta p$ and $\Delta x$ according to Eqns.~(\ref{eqns:deltas}) and update $(x,p)$ to correspond to this choice.
    \item Simulate the spontaneous emission process further by updating $p$ to account for the momentum recoil by selecting one of the three possibilities in the discretely represented dipole radiation pattern. Assign ${\ket{\psi(t_j)} = \ket{\widetilde{0}}}$.
    \item Relabel the basis subset $W(p_r)$ with the updated momentum $p$. 
    \item Replace $\ket{\psi_r},t_r$ and $(x_r,p_r)$ with their updated values $\ket{\widetilde{0}}$, $t_j$, and $(x,p)$.
    \end{enumerate}
    \end{description}
    \item \label{reloop} Repeat as needed from step \ref{evolve} until a desired final time is reached.
\end{enumerate}
We choose to perform the operations described in Case~A at the middle and end of each sweep because the particle ideally undergoes one adiabatic transfer per half sweep by the design of the SWAP procedure, and hence experiences the impulse of a single photon momentum.

As a framework for comparison and consistency, we have developed an analogous model for a traditional 1D MOT, which we use in Section~\ref{phasespace}. In that case, the phase space variables are only updated whenever a spontaneous event occurs, as this is typically the only mechanism whereby the system can traverse momentum space. The resulting cloud size, temperature, and spontaneous photon record are in close agreement with other models~\cite{metcalf}. Moreover, the SWAP MOT algorithm we present agrees well with a fully quantum $\sigma^+-\sigma^-$ SWAP cooling model with no magnetic trapping (as opposed to the configuration previously developed in \cite{exp,swap_theory} with linearly polarized lasers), which can easily be implemented computationally due to its translational invariance.


\section{Phase Space Dynamics} \label{phasespace}

\begin{figure}
    \centering
    \includegraphics[width=\linewidth]{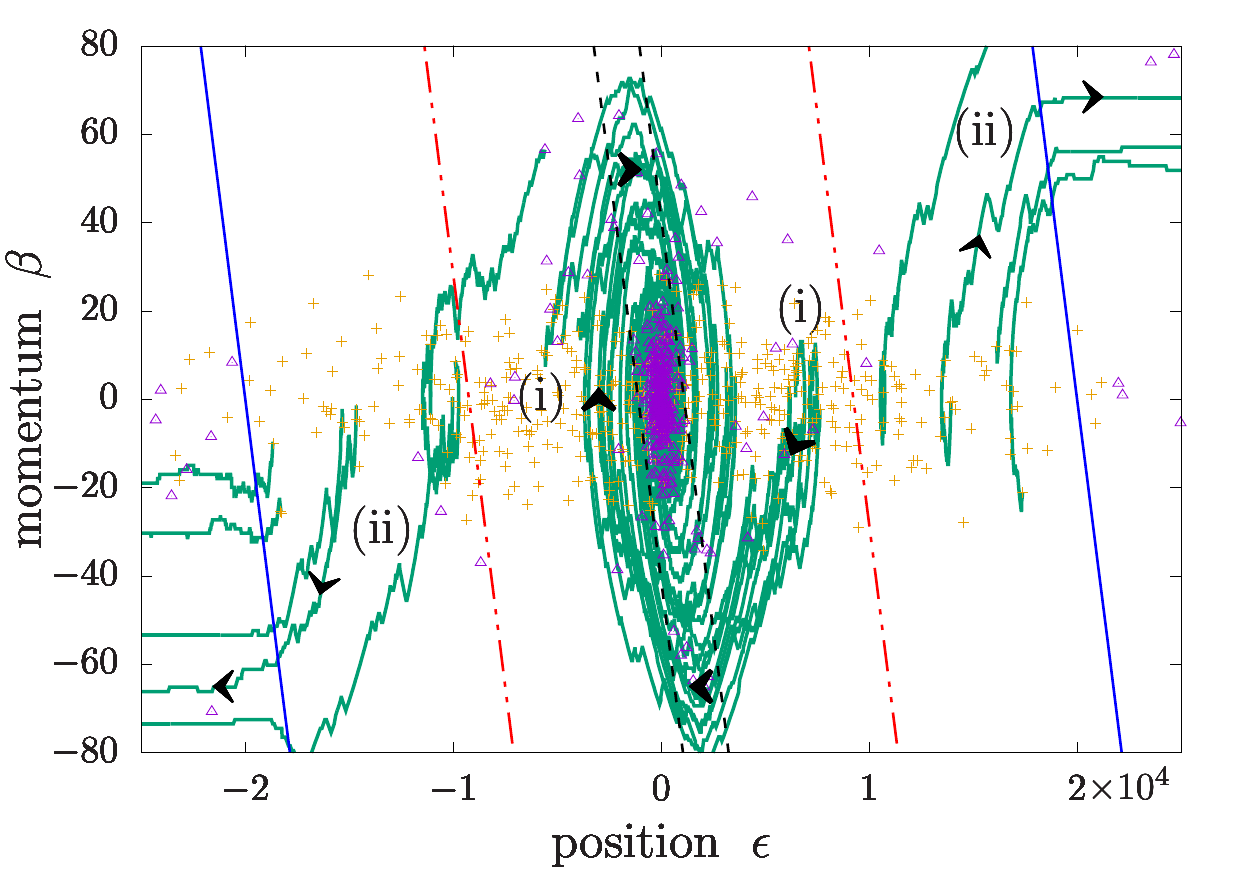}
    \caption{Phase space dynamics in a SWAP MOT. The variables $\beta$ and $\epsilon$ label a particle's momentum as a multiple of the number of photon momenta and number of inverse wavenumbers from the center of the trap, respectively. The diagonal lines, which have slopes determined by Eq.~(\ref{eq:psslope}), separate phase space into regions that are labeled above the plot and characterized by Eq.~(\ref{eq:psregions}). Several sample simulated phase space trajectories are displayed as green lines. The black arrows show the flow of time. Particles in the ``ideal" region, labeled (i), are generally cooled and trapped, while those outside the ``ideal" region, labeled (ii), are heated out of the trap. Each of the particles' initial momenta and positions (yellow targets) were sampled from Gaussian distributions with standard deviations $\sigma_{\beta_0} = 11.5$ and $\sigma_{\epsilon_0} = 9.1 \cdot 10^3$, respectively. Parameters, in units of $\gamma$, are: $\omega_r = 0.67,  M=0.05, \Omega = 55, \Delta_s = 2000, T_s = 1.5,$ and $t_\text{max} = 375.$ For this set of parameters, the particles are slightly cooled and become significantly more trapped (final coordinates are displayed as purple triangles).}
    \label{fig:traj}
\end{figure}

\noindent Now that we have developed a detailed semiclassical model, we are in a position to simulate and characterize the dynamics in a SWAP MOT. In this section, we first provide general insight for appropriate choices of the many experimental parameters. Then, as a relevant physical example, we simulate the dynamics of the SWAP MOT procedure as applied to the molecule yttrium monoxide (YO) using parameters for one of its narrow linewidth transitions. We use the simulation results to further elucidate the dynamics in various regions of phase space, which are separated by diagonal lines with slopes determined by the motional detuning (see Fig.~\ref{fig:traj}):
\begin{equation}
\label{eq:psslope}
    \delta^\text{m} = bx + kv = M \epsilon + 2 \beta \omega_r \; \Rightarrow \;
    \frac{d\beta}{d\epsilon} = - \frac{M}{2 \omega_r}.
\end{equation}
Here, $M \equiv b/k$ is the frequency that characterizes the magnetic field gradient, and $\epsilon \equiv kx$ and $\beta \equiv p/\hbar k$ are the position and momentum expressed in the appropriate dimensionless units corresponding to the optical transition. Finally, we compare the results of the trap loading process and phase space compression to those in a traditional MOT.

\subsection{Choosing appropriate experimental parameters}
\label{conditions}

\noindent The essential constraints  for the experimental parameters arise from the requirement of rapid adiabatic passage, which forms the foundation of the SWAP method. Landau and Zener proved~\cite{zener} that population is efficiently transferred between two stable quantum states as long as the laser frequency is linearly swept over the energy splitting of the states with the following condition satisfied:
\begin{equation}
\label{eq:adiabaticity}
    \kappa \equiv \frac{\Omega^2}{\alpha} > 1,
\end{equation}
where we define $\kappa$ as the adiabaticity parameter and $\alpha~\equiv~\Delta_s / T_s$ is the slope of the frequency ramp~\cite{zener}. Experimental parameters that satisfy Eq.~(\ref{eq:adiabaticity}) are said to operate within the adiabatic regime. 

Additionally, it can be shown \cite{vitanov} that the transfer process takes a time $\tau_\text{j}=2\Omega/\alpha$ in the adiabatic regime. Because it is necessary for the system to remain in the excited state without spontaneously decaying for at least a time $\tau_\text{j}$, it must satisfy $\tau_\text{j} \ll 1/\gamma $. Combining this result with Eq.~(\ref{eq:adiabaticity}), we find that the Rabi frequency must satisfy
\begin{equation}
\label{eq:rapid}
    \Omega \gg \gamma.
\end{equation}
This is exactly what is meant by adiabatic {\it rapid} passage \cite{abragam}.

We now discuss the choice of the sweep range~$\Delta_s$. In Section~\ref{mechanism}, we described how a particle can experience a coherent force and a consequent impulse of $2 \hbar k$ per sweep if it begins in the internal ground state $\ket{0}$. However, it will still experience an impulse that leads to cooling and trapping, albeit smaller in magnitude, even if it begins with some excited state population. By symmetry, this impulse will aid in cooling and trapping a particle only if its wave function predicts that it is more likely to begin a sweep in $\ket{0}$. One way to guarantee this condition is to enforce an imbalance such that the particle spends more time outside the two photon resonances, allowing spontaneous emission to preferentially optically pump the system to $\ket{0}$ for the next sweep. The most obvious way to do this is to introduce a waiting period in between each sawtooth ramp. However, because it is often desirable to cool and trap as quickly as possible, we only consider situations with no waiting period. In this case, we achieve a time imbalance whenever
\begin{equation}
\label{eq:sweepRestriction}
4 |\delta^\text{m}| \leq \Delta_s,
\end{equation}
since the resonances are spaced in frequency by $2|\delta^\text{m}|$. 

Next, we consider restrictions on the sweep period~$T_s$. In the sequence discussed in Section~\ref{mechanism}, the particle spends a time $\tau_\text{e} \approx 2 |\delta^\text{m}|/\alpha$ in one of the excited states in between the two resonances. (This can be seen by writing the detuning during a single sweep as $\delta(t)=\alpha t$ and noting that the lasers resonate with the particle when $\delta(t) = \pm \delta^\text{m}$.) Therefore, in order for the particle to maintain a low probability of spontaneous emission in between resonances, $\tau_\text{e}$ must satisfy
\begin{equation}
\label{eq:excitedTime}
    \tau_\text{e} \approx \frac{2 |\delta^\text{m}|}{\alpha} < 1/\gamma.
    \quad \Rightarrow \quad
    4 |\delta^\text{m}| < \frac{2 \Delta_s}{\gamma T_s }.
\end{equation}
The resetting to the internal ground state enforced by the time imbalance is only useful if the particle does not have a significant chance of decaying between resonances during the next sweep. So, applying Eq.~(\ref{eq:sweepRestriction}) and Eq.~(\ref{eq:excitedTime}), a reasonable restriction on the sweep period $T_s$ is given by
\begin{equation}
\label{eq:tRestriction}
    \gamma T_s < 2\,,
\end{equation}
in order to obtain fast results. It has been shown experimentally that a cooling force twice as large as the radiation pressure force was achieved in a ${}^{163}$Dy system with the choice $\gamma T_s \approx 0.1$ \cite{dysprosium}. However, some experiments may require the much stricter condition of minimal spontaneous events. In these cases, one should instead come close to saturating the bound in Eq.~(\ref{eq:tRestriction}) so that there is substantial time to reset to the ground state via spontaneous emission in the event of incorrect time-ordering of absorption and emission.

We now derive a condition for the choice of magnetic field gradient. As will be explained in Section~\ref{ideal}, there is a specific region of phase space, which we call the ``ideal region," where particles generally undergo a change in momentum of $2 \hbar k$ per sweep via the coherent dynamics discussed in Section~\ref{mechanism}. Here, it is a good approximation to expand the momentum to first order in time, i.e.,
\begin{align}
\label{eq:beta(t)}
    \beta(\tau) &\approx \beta_0 + \frac{d\beta}{d \tau} \tau \notag \\
    & \approx \beta_0 - 2 \text{sgn}(\delta^\text{m}_0) \tau.
\end{align}

\noindent In Eq.~(\ref{eq:beta(t)}), $\tau \equiv t/T_s$ labels the number of sweeps, and we have made the approximation that the momentum transfer $\Delta \beta =2$ is uniformly distributed over a sweep instead of localized near photon resonances. Note that the sign function arises because the sign of $\delta^\text{m}$ determines which laser the system interacts with first, and therefore the direction of the impulse. This family of trajectories has an instantaneous slope of magnitude 
\begin{equation}
\label{eq:slope}
    \left|
    \frac{d \beta}{d \epsilon}(\tau)
    \right|  =
    \frac{1}{\omega_r T_s|\beta_0 - 2\text{sgn}(\delta^\text{m}_0) \tau|} \leq \frac{1}{\omega_r T_s | \beta_0|},
\end{equation}
which was derived by applying the integral relationship $x = \int v \, dt$, equivalent to $\epsilon=2 \omega_r T_s \int \beta \, d\tau$. Any trajectory with a slope whose magnitude is smaller than Eq.~(\ref{eq:psslope}) has a substantial likelihood of moving out of the ideal region and into an undesirable ``heating region," described in Section~\ref{heating}. Therefore, it is reasonable to bound the slope of our trajectories, as given in Eq.~(\ref{eq:slope}), from below by the the slope of the diagonal lines that define different phase space regions [Eq.~(\ref{eq:psslope})], which leads to the condition
\begin{equation}
\label{eq:mrestriction}
    M T_s \leq 2/|\beta_0|.
\end{equation}
Because the minimum achievable temperature in SWAP cooling is the recoil temperature \cite{swap_theory}, it is safe to assume that the momenta of most of the particles satisfy $|\beta_0|>1$. 

As mentioned earlier, it is necessary to choose a sweep period $T_s$ that comes close to saturating Eq.~(\ref{eq:tRestriction}) when it is a priority to minimize spontaneous events. Combining Eqns.~(\ref{eq:tRestriction}) and~(\ref{eq:mrestriction}) in this limit yields the restriction 
\begin{equation}
\label{eq:mLg}
    M \leq \gamma.
\end{equation} 
Together, Eqns.~(\ref{eq:rapid}) and (\ref{eq:mLg}) motivate our omission of the coherent Zeeman force in Section~\ref{model}.

\subsection{Cooling and trapping of the molecule YO}
\label{YOsim}

\noindent Figure~\ref{fig:traj} displays simulated phase space trajectories for a cloud of Yttrium Monoxide (YO) particles being cooled and trapped in a SWAP MOT using its 5.9 kHz linewidth ${{X}^{2}}\Sigma^+ \rightarrow A{{^{\prime} }^{2}}{{\Delta }_{3/2}}$ transition~\cite{yoparameters}.
Motivated by recent experimental progress~\cite{yoparameters}, the initial particle distribution was characterized by Gaussians with standard deviations $\sigma_{\epsilon_0}= 9.1 \cdot 10^3$ and~$\sigma_{\beta_0}=11.5$, and is displayed as yellow target symbols ``$+$" in Fig.~\ref{fig:traj}. By comparing the initial and final particle distributions, the latter displayed as purple triangles, it is clear that phase space compression was achieved. We quantify this result in Section~\ref{compression}.

The simulation parameters were chosen under the conditions derived in Section~\ref{conditions}. 
Because we wish to highlight the SWAP MOT's ability to cool and trap with fewer scattered photons, the sweep period was chosen to satisfy $\gamma T_s = 1.5$, which is close to the bound in Eq.~(\ref{eq:tRestriction}).
The magnetic field gradient was consequently chosen to be $M=0.05\gamma$ in order to yield significant trapping with a capture range roughly corresponding to any particles with momenta smaller in magnitude than $3 \sigma_{\beta_0}$ [see Eq.~(\ref{eq:mrestriction})].
These choices bounded the motional detuning of most particles to the interval
\begin{equation}
    |\delta^\text{m}| \leq M \sigma_{\epsilon_0} + 2 \sigma_{\beta_0}\omega_r  \approx 470 \gamma.
\end{equation}
Therefore, the choice $\Delta_s = 2000 \gamma$ guaranteed that most particles underwent the dynamics described in Section~\ref{mechanism} [Eq.~(\ref{eq:sweepRestriction})] and did not frequently emit spontaneous photons in between the two resonances [Eq.~(\ref{eq:excitedTime})].
Finally, to operate in the regime of adiabatic rapid passage, the Rabi frequency was chosen to be $\Omega = 55 \gamma$ [see Eqns.~(\ref{eq:adiabaticity}) and (\ref{eq:rapid})]. 

\subsection{Characteristic dynamics in different phase space regions}

\noindent  The diagonal lines in Fig.~\ref{fig:traj} divide phase space into regions with different characteristic behavior, which we now discuss in the following sections.

\subsubsection{Region of ideal coherent dynamics}
\label{ideal}

\noindent Upon investigation of the coherent dynamics under Eq.~(\ref{eq:SCH}) in the adiabatic limit, we can deduce that a particle initialized in the internal ground state $\ket{0}$ populates a single excited state during a sweep only if $|\delta^\text{m}| >|\Omega|$. (In our previous work, we referred to a similar condition as the ``high-velocity regime" \cite{swap_theory}.) Otherwise, power broadening would bring both lasers into resonance with the particle simultaneously, distributing population among all three states. Specifically, a particle with $\delta^\text{m}>|\Omega|$ ($\delta^\text{m}< -|\Omega|$) first absorbs a photon from the $\sigma^-$ ($\sigma^+$) laser. Then, after the laser polarizations and the magnetic field are abruptly switched, the particle eventually emits a photon into the other laser, which now has the correct polarization (see Fig.~\ref{fig:energy}). Thus, the particle experiences a negative (positive) impulse of magnitude $2 \hbar k$ after each sweep in the absence of spontaneous emission. 

When spontaneous emission is included, the particle must not spend too much in the excited state $\tau_\text{e}$ [see Eq.~(\ref{eq:excitedTime})] or else dissipation will occur. In order to retain mostly coherent dynamics, we propose that $\tau_\text{e}$ must be smaller than the time $\log(2)/\gamma$, as half of the particles will decay on average before resonating with the other laser in this time. Combining these results, we expect coherent dynamics to approximately describe any particles with motional detunings $\delta^\text{m}$ that satisfy  
\begin{equation}
\label{eq:idealDynamics}
    |\Omega| < |\delta^\text{m}| < \alpha \log(2)/2 \gamma.
\end{equation} 

\noindent We will call this the ``ideal" region, as most particles that are initialized in this region are eventually trapped and cooled. As seen in Fig.~\ref{fig:traj}, particles in the ideal region, which are labeled (i), follow trajectories that generally exhibit an effective ``attraction" to the lines $\delta^\text{m} = \pm |\Omega|$. It should be noted that the black arrows display the direction of the flow of time during the simulations.

\subsubsection{Dynamics near $\delta^\text{m}=0$}

\noindent In this regime, the Zeeman and Doppler shifts are comparable in magnitude but opposite in sign. Thus, the resonances with each laser are no longer time resolved, resulting in population of both excited states and therefore frequent spontaneous emission events. Nevertheless, we observe an average attraction of the trajectories to the line $\delta^\text{m} = 0$ in phase space. This attraction paired with ballistic expansion causes the particle to eventually drift toward the phase space origin in a similar manner to the dynamics within a traditional MOT. After many cycles, the system equilibrates about the phase space origin, as discussed in Section \ref{equilibrium}.

\subsubsection{Dynamics outside the ideal region}
\label{heating}

\noindent The condition $|\delta^\text{m}| < \Delta_s/2$ is necessary to ensure that the particle resonates with both lasers at some point during the sweep. Obviously, particles outside of this regime will undergo purely ballistic expansion, as seen by the trajectories labeled (ii) at later times in Fig.~\ref{fig:traj}.

If a particle is in the region defined by $\alpha \log(2)/2 \gamma < |\delta^\text{m}|< \Delta_s/2$, there is a significant chance of spontaneous emission before resonating with the second laser in a sweep. If the spontaneous emission does occur, the particle will absorb a photon when it resonates with the second laser, directing it even further from the ideal region. In this sense, the particle is heated and will eventually escape the trap, as evidenced by the trajectories labeled (ii) at earlier times in Fig.~\ref{fig:traj}. The most extreme case of heating occurs when $|\delta^\text{m}| \approx \Delta_s/2$; here, the absorption-emission sequence becomes centered at the extremes of the sawtooth wave instead of at the center, leading to coherent transfer of the particle away from the phase space origin.

Assembling these considerations together, we arrive at the following complete picture of the phase space regions, labeled in Fig.~\ref{fig:traj}:
\begin{align}
\label{eq:psregions}
    |\delta^\text{m}| \leq |\Omega| \qquad &(\text{overlap, scattering}) \notag \\
    |\Omega| < |\delta^\text{m}| < 
      \frac{\alpha \log(2)}{2 \gamma} \qquad
      &(\text{ideal coherent dynamics}) \notag \\
          \frac{\alpha \log(2)}{2 \gamma}
      < |\delta^\text{m}| 
      < \frac{\Delta_s}{2} \qquad
      &(\text{heating}) \notag \\
    \frac{\Delta_s}{2}    
      < |\delta^\text{m}| \qquad
      &(\text{ballistic expansion}).
\end{align}

\begin{figure*}
    \centering
    \includegraphics[width=0.9\linewidth]{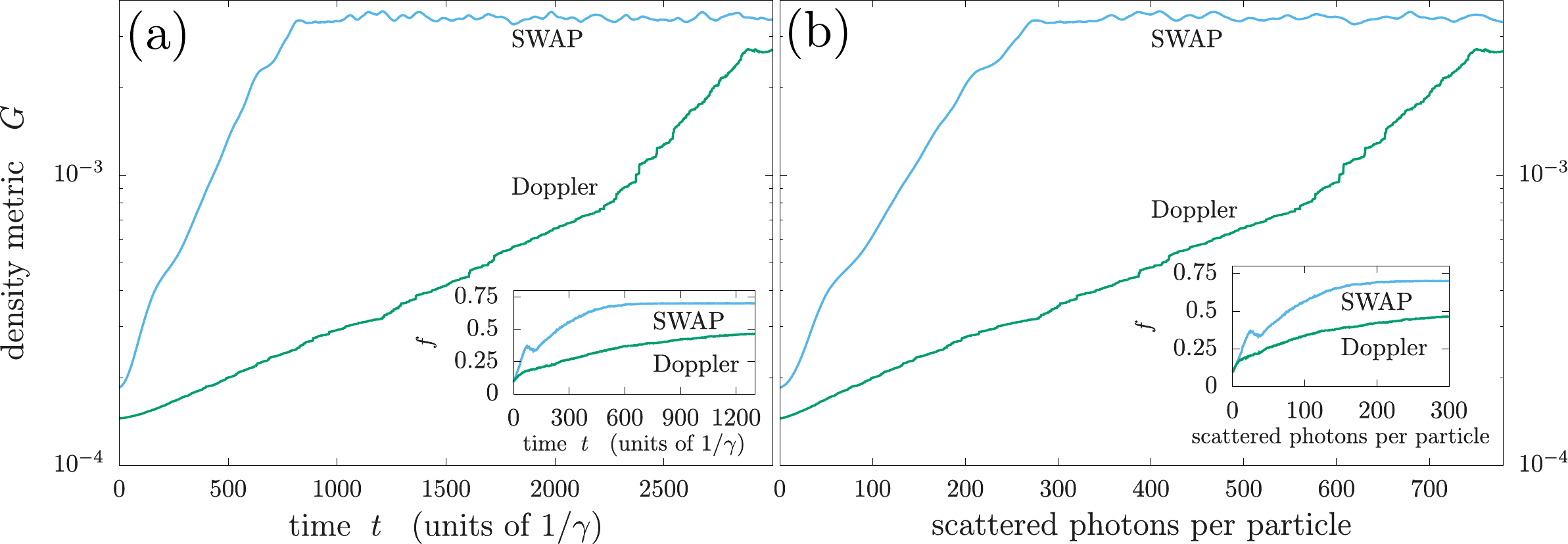}
    \caption{Semilog plots of the evolution of the phase space density metric $G$ for the YO system subject to the SWAP MOT and traditional MOT (Doppler). The used SWAP MOT simulation parameters were identical to those in Fig.~\ref{fig:traj}. The detunings of the lasers in the traditional MOT simulation were set to $\delta = - \Omega$, and all other parameters were identical to the SWAP MOT simulation. Quantities were averaged over all trajectories that become trapped within the overlap region [see Eq.~(\ref{eq:psregions})] from $N_\text{tot}=2000$ total trajectories. (a) $G$ versus time $t$ for all trapped particles. The SWAP MOT equilibrates more quickly than the traditional MOT. Inset: The SWAP MOT captures a higher fraction $f(t)$ of particles. (b) $G$ versus the average number of spontaneous emissions per particle. Spontaneous emissions of non-trapped particles are not included. The SWAP MOT equilibrates with significantly fewer spontaneous emissions per trapped particle.}
    \label{fig:psd}
\end{figure*}

\noindent In summary, particles that begin in the ideal region are transferred to the overlap region with few scattering events. Once there, the particle frequently scatters photons and the system eventually equilibrates. Any particles that begin outside the ideal region are most likely heated out of the trap.

\subsection{Phase space compression}
\label{compression}

\noindent The ultimate signature of a laser cooling method's ability to effectively trap and cool a system is the ability to overcome the limits imposed by Liouville's theorem for Hamiltonian evolution and exhibit compression in terms of occupied phase space volume. One useful metric that can be used to quantify the particle density in phase space is the root-mean-square (rms) interparticle spacing in phase space, $l_\text{rms}$. If we label the position of a particle in phase space with the dimensionless vector $\bs{r} \equiv \{\epsilon, \beta\}$, then $l_\text{rms}$ is
\begin{equation}
\label{eq:lbar}
    l_\text{rms} \equiv \frac{1}{N}
        \sqrt{
            \sum_{i>j}^{N}
            \left|
                \bs{r}_i-\bs{r}_j
            \right|^2
        },
\end{equation}

\noindent where $N$ is the number of trajectories considered out of the total number $N_\text{tot}$. For our purposes, we only need to consider the trajectories that ultimately become trapped within the overlap region. We define a metric $G$ for the particle phase space density as 
\begin{equation}
    G \equiv 1/l_\text{rms},
\end{equation} 

\noindent and we label the fraction of trapped particles with $f = N/N_\text{tot}$.

In order to gain intuition into the interpretation of $G$, let us briefly consider its form for a system with a phase space distribution $F(\bs{r})$ that can be modeled as a product of Gaussian distributions in $\epsilon$ and $\beta$ with standard deviations $\sigma_\epsilon$ and $\sigma_\beta$:
\begin{equation}
\label{eq:gaussianPSD}
F(\bs{r}) =
    \frac{N}{2 \pi \sigma_\epsilon \sigma_\beta}
    \exp \left[
        - \frac{1}{2} \left(
            \frac{\epsilon^2}{\sigma_\epsilon^2}
            +\frac{\beta^2}{\sigma_\beta^2}
        \right)
    \right].
\end{equation}
The continuous form of Eq.~(\ref{eq:lbar}) is
\begin{equation}
\label{eq:lbarcont}
    l_\text{rms} =
    \frac{1}{N} \sqrt{
        \frac{1}{2}
        \int dr_1 \, dr_2 \, F(\bs{r}_1) F(\bs{r}_2)
            |\bs{r}_1 - \bs{r}_2|^2
        },
\end{equation}
so using Eq.~(\ref{eq:gaussianPSD}) in Eq.~(\ref{eq:lbarcont}) results in
\begin{equation}
    G = 1/\sqrt{\sigma_\epsilon^2+\sigma_\beta^2}.
\end{equation} 
Further, let us consider the case when the system is in the Bose-Einstein condensate (BEC) regime. Here, the system would occupy a single volume element of phase space, which corresponds to $\Delta x \Delta p / h = \Delta \epsilon \Delta \beta/2 \pi \approx 1$. In the symmetric case $\Delta \epsilon = \Delta \beta$,  we find that $G$ is on the order of unity. These results help with interpreting the resulting scale of the calculated values for $G$ in the SWAP MOT, where $G$ will always be considerably below the quantum degenerate value.

Figures \ref{fig:psd}(a) and \ref{fig:psd}(b) show the evolution of $G$ and $f$ for the YO transition introduced in Section~\ref{YOsim} subject to both a traditional MOT and the SWAP MOT as a function of time along with the average number of spontaneous emissions per particle. Only particles that ultimately settle within the overlap region [see Eq.~(\ref{eq:psregions})] are considered. The insets display the fraction $f$ of particles that are currently within the overlap region. It is evident that both procedures compressed phase space, but the SWAP MOT did so more than three times faster and with roughly a third of the spontaneous emissions required by the traditional MOT for this set of parameters. This result shows that the SWAP MOT can apply higher dissipative forces and remove entropy with a higher photon efficiency, i.e., it removes more energy and momentum per scattered photon than a traditional MOT. 

\begin{figure}
    \centering
    \includegraphics[width=\linewidth]{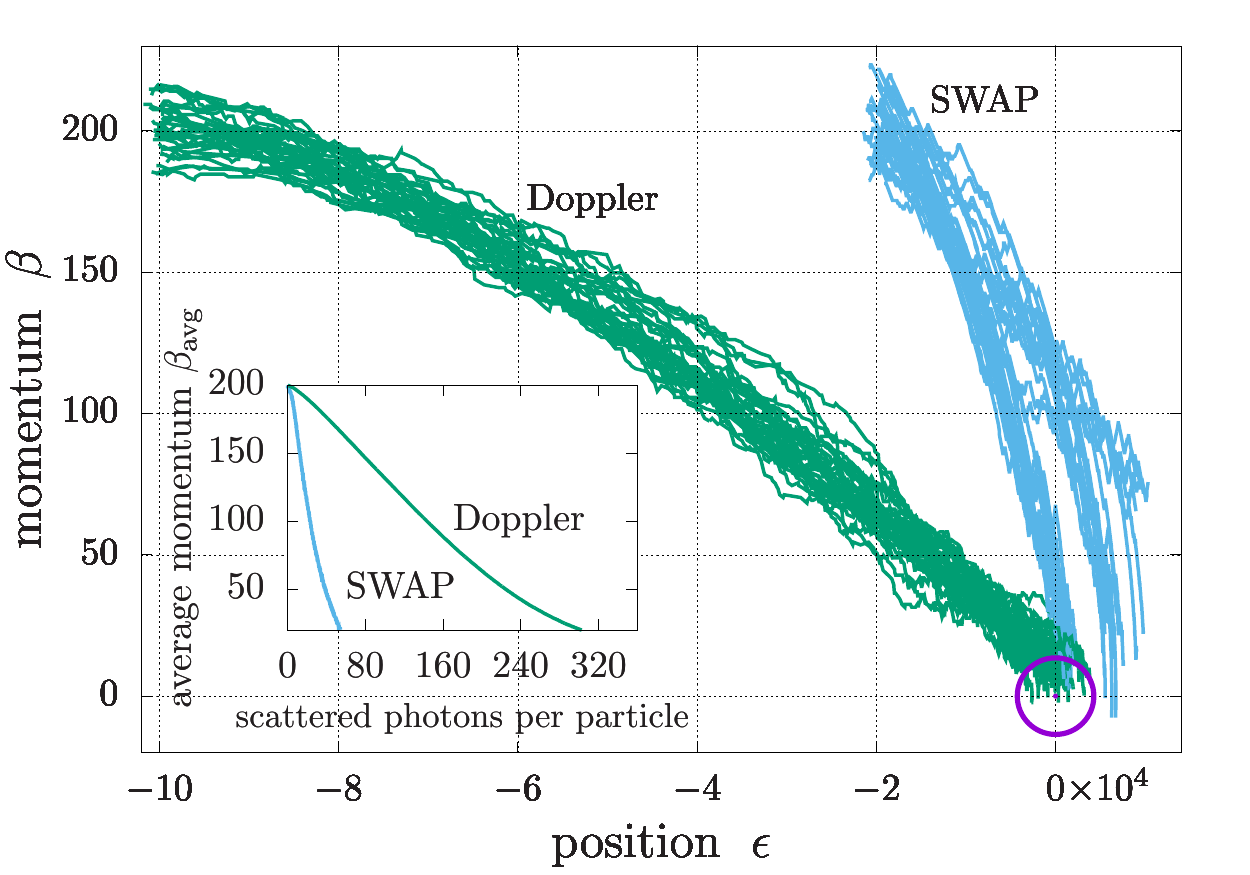}
    \caption{Simulated phase space trajectories of particles being loaded into a SWAP MOT and traditional MOT (Doppler). The trajectories generally flow toward the phase space origin, which is labeled by a purple circle. The particles subject to the SWAP MOT experience much higher forces compared to the radiation pressure force intrinsic to the traditional MOT. Common parameters, in units of $\gamma: \omega_r =0.67, \Omega = 60, \beta_0=200, \sigma_{\epsilon_0} =1000, \sigma_{\beta_0}=10$. SWAP MOT parameters: $M=0.012,\Delta_s=1600, T_s=1, t_\text{max}=175, \epsilon_0=-2 \cdot 10^4.$ Traditional MOT parameters: $M=0.0025, \delta=-60,t_\text{max}=1000, \epsilon_0=-10^5.$ Inset: The average momentum number $\beta_\text{avg}$ versus the average number of spontaneous emissions per particle. The SWAP MOT is loaded with fewer spontaneous photons and in a shorter distance.}
    \label{fig:load}
\end{figure}

\subsection{MOT loading} \label{load}

\noindent The previous subsection showcases the SWAP MOT's ability to efficiently trap and cool. Here, we demonstrate its ability to coherently translate high momentum states toward zero momentum, which is desirable in, e.g., the loading of molecules from an oven or supersonic nozzle for which a low number of scattered photons is a priority. 

Commonly utilized particle loading devices, such as the Zeeman slower, rely on radiation pressure to reduce the particle speed and yield a force that is fundamentally limited by the linewidth of the transition~\cite{metcalfText}. Consequently, such a process requires roughly one spontaneous photon per absorbed slowing photon. In contrast, the SWAP protocol is fundamentally limited by the rate at which one can can stimulate rapid adiabatic passage. Hence, operating in the regime $\Omega \gg \gamma$ yields significantly higher coherent forces and therefore smaller slowing distances with fewer scattering events. 

Figure~\ref{fig:load} shows the results of simulating the slowing of YO on the 5.9 kHz transition using both a traditional MOT and a SWAP MOT. We chose the Rabi frequency $\Omega = 60 \gamma$ for both procedures to allow for high coherent forces in the SWAP MOT and rapid absorption in the traditional MOT. The laser detunings in the traditional MOT were set to $\delta = -\Omega$ so that the particles were close to resonance with the counterpropagating laser at the beginning of the simulation. The SWAP magnetic field gradient $M=0.012 \gamma$ was chosen so that the trajectories lay on the border between the overlap and ideal regions for the entire slowing process, which minimized the amount of time the particles spent in the excited states while keeping the resonances sufficiently separated. It is evident that this procedure applied forces to the particles in the SWAP MOT that were roughly five times larger than the traditional MOT and consequently slowed them in roughly one-fifth of the distance. As shown in the inset of Fig.~\ref{fig:load}, the SWAP procedure produced roughly a fifth of the spontaneous photons required by the traditional MOT in order to reach the region of the phase space origin for this set of parameters. Therefore, the SWAP MOT may potentially be a strong candidate for molecular slowing, as it produces high coherent forces with few spontaneous photons, which may thereby reduce the adverse effect of optical pumping into dark states.


\section{equilibrium temperatures and cloud sizes} \label{equilibrium}

\noindent The simulated phase space diagram shown in Fig.~\ref{fig:traj} and the resulting phase space densities shown in Fig.~\ref{fig:psd} suggest that particles captured in a SWAP MOT eventually equilibrate to a final, fixed phase space distribution. In this section, we describe the associated equilibrium momentum and position distributions by analyzing their dependence on both laser intensity and the ratio $\omega_r/\gamma$, where the latter quantity characterizes the separation between the external and internal timescales.

\begin{figure}
    \centering
    \includegraphics[width=0.9\linewidth]{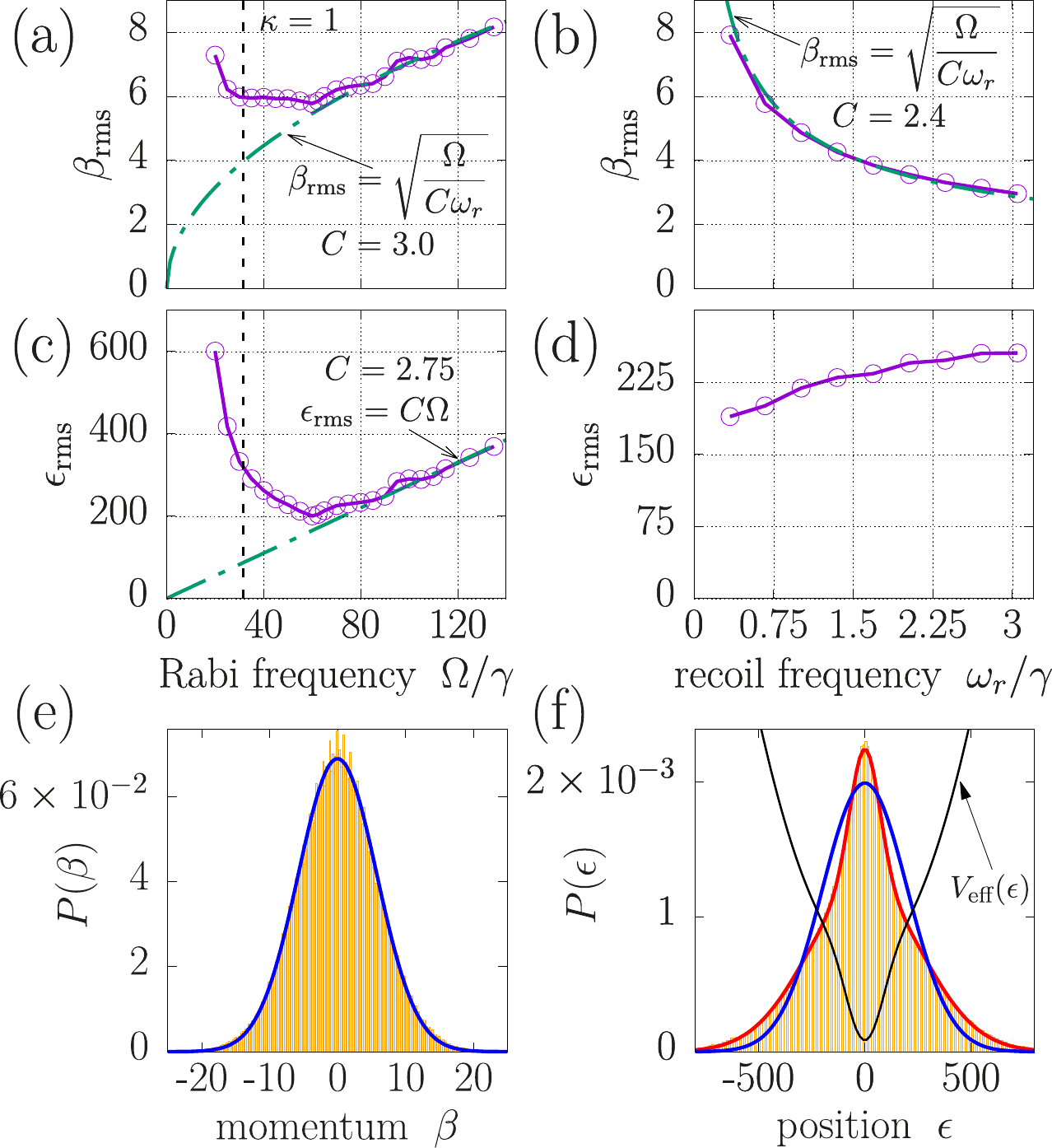}
	\caption{(a-d): Equilibrium root mean square momenta $\beta_\text{rms}$ and positions $\epsilon_\text{rms}$ as a function of Rabi frequency $\Omega$ [(a) and (c)] and recoil frequency $\omega_r$ [(b) and (d)], both in units of the linewidth $\gamma$. The vertical, dotted black lines in subplots (a) and (c) display the Rabi frequency for which the adiabaticity parameter $\kappa=1$ [see Eq.~(\ref{eq:adiabaticity})]. Numerical fits to plots (a-c), which are displayed as green dot-dashed curves, are discussed in the text. Common parameters, in units of $\gamma$, are: $M=0.05, \Delta_s = 2000,$ and $T_s=2$. The recoil frequency in plots (a), (c), (e) and (f) is $\omega_r = 0.67\gamma$, and the Rabi frequency in plots (b), (d), (e) and (f) is $\Omega=60\gamma$. Each point (purple circles) was averaged over data taken from 800 independent trajectories, each being sampled once per sweep over 1,500 sweeps. (e-f): Equilibrium momentum and position probability distributions $P(\beta)$ and $P(\epsilon)$. Gaussian fits (blue curves), which were used to extract root-mean-square values, are discussed further in the text. Histogram data was taken from 800 independent trajectories, each being sampled once per sweep over 1,500 sweeps. In (f), the black curve is the effective potential $V_\text{eff}(\epsilon)$ [Eq.~(\ref{eq:vEFf})] that generates the non-Gaussian position distribution [Eq.~(\ref{eq:positionDist}), red curve].}
    \label{fig:eq}
\end{figure}

The equilibrium momentum and position probability distributions $P(\beta)$ and $P(\epsilon)$ from a YO SWAP MOT simulation are displayed in Figs.~\ref{fig:eq}(e) and~\ref{fig:eq}(f). The parameters were identical to those in Fig.~\ref{fig:traj}, except $\Omega=60\gamma$ here. Both distributions can approximately be described by a normalized Gaussian function with zero mean, as shown by the fits displayed as blue curves. In this way, we can consider the measured width to the momentum distribution as approximately the standard deviation of a thermal 1D Maxwellian distribution. From the equipartition theorem,
\begin{equation}
\label{eq:equipartition}
  \frac{p_\text{rms}^2}{2m}
    =\hbar \omega_r \beta_\text{rms}^2
    = \frac{1}{2} k_B T,
\end{equation}
thereby defining an effective temperature $T$. We find this type of fit to be much more accurate for the momentum distribution than is observed for the position distribution. Nevertheless, the points displayed as purple circles in Figs.~\ref{fig:eq}(a-d) are extracted by fitting each resulting probability distribution to a Gaussian with zero mean.

Figures~\ref{fig:eq}(a) and~\ref{fig:eq}(c) display the equilibrium root-mean-square momenta $\beta_\text{rms}$ and positions $\epsilon_\text{rms}$ for various values of $\Omega/\gamma$ using the YO transition introduced in Section~\ref{YOsim}. These plots provide information about the effects of the laser intensity $I$ on the equilibrium phase space distribution, since $\Omega/\gamma \propto \sqrt{I}$ \cite{metcalf}. One reason these plots are of interest is because the lowest temperature in other cooling methods is typically achieved in the limit of low saturation, i.e., $\Omega \rightarrow 0$ (although in this limit the cooling may take a long time and the trapping effect is reduced), which is not usually experimentally reasonable due to the demands of adiabaticity [Eq.~(\ref{eq:adiabaticity})] and rapid adiabatic passage [Eq.~(\ref{eq:rapid})]. The vertical, black dashed line separates the plots into diabatic (left) and adiabatic (right) regimes. We observe that the minimum temperature and cloud size is achieved slightly within the adiabatic region, where particles are efficiently transferred into resonant states with minimal negative effects from power broadening. A similar result for the minimum temperature was previously discovered for SWAP cooling \cite{exp,swap_theory}.

Figures~\ref{fig:eq}(b) and~\ref{fig:eq}(d) display the equilibrium root-mean-square momenta $\beta_\text{rms}$ and positions $\epsilon_\text{rms}$ for various values of $\omega_r/\gamma$ when $\Omega = 60 \gamma$. This ratio is a fundamental property of the cooling transition, so this investigation transcends our focus on the specific YO transition used elsewhere in this work and provides a better understanding of the SWAP protocol's efficacy on an arbitrary transition. As seen in Fig.~\ref{fig:eq}(b), $\beta_\text{rms}$ decreases as $\omega_r/\gamma$ increases, which provides direct evidence for our claim that SWAP procedures applied to narrow linewidth transitions result in lower temperatures $T$. On the other hand, we observe an increase in cloud size in  Fig.~\ref{fig:eq}(d).

In the deeply adiabatic regime, we observe that the root-mean-square values scale as
\begin{equation}
\label{eq:Eqtrends}
    \beta_\text{rms} \propto 
        \sqrt{\frac{\Omega}{\omega_r}};
    \qquad
    \epsilon_\text{rms} \propto 
        \Omega.
\end{equation}
Numerical fits according to the equation
\begin{equation}
	\beta_\text{rms} = \sqrt{\frac{\Omega}{C\omega_r}}
\end{equation}
are displayed as green dot-dashed curves in Figs.~\ref{fig:eq}(a) and \ref{fig:eq}(b). Figure~\ref{fig:eq}(c) is numerically fit to the equation
\begin{equation}
	\epsilon_\text{rms} = C \Omega.
\end{equation}
The scaling constants $C$ for each independent fit are displayed within each plot. Combining Eq.~(\ref{eq:equipartition}) with the observed trend in Eq.~(\ref{eq:Eqtrends}), we find that
\begin{equation}
\label{eq:swapMotTemp}
    k_B T \propto \hbar \Omega,
\end{equation}

\noindent which agrees with previous studies of SWAP cooling \cite{swap_theory}.

As shown by the red curve in Fig.~\ref{fig:eq}(f), the position distribution is actually more accurately described by the sum of two Gaussians of different widths. This suggests that the effective potential $V_\text{eff}(x)$ describing both the conservative and dissipative effects of the SWAP MOT has an exaggerated dip near $x=0$, similar to that of a dimple trap \cite{dimple}. As shown in Appendix \ref{dimple}, $V_\text{eff}(x)$ can be derived from the resulting equilibrium position distribution, and we include it as a black curve in Fig.~\ref{fig:eq}(f).

In a traditional MOT, the equilibrium temperature $T_\text{trad}$ scales with the linewidth $\gamma$ \cite{metcalf}, {i.e.},
\begin{equation}
    k_B T_\text{trad} \propto \hbar \gamma.
\end{equation}
providing this is larger than the recoil temperature.
Comparing this to Eq.~(\ref{eq:swapMotTemp}), we find that the temperature reached in a SWAP MOT must be larger since $\Omega \gg \gamma$. As was discussed in the original SWAP cooling papers \cite{exp,swap_theory}, we again reach the conclusion that the utility of the SWAP MOT is not its low temperatures, but its ability to apply larger forces and reach equilibrium faster and with fewer scattered photons. This is especially applicable to systems with narrow line transitions, i.e., small~$\gamma$.

\section{Conclusion}\label{conclusion}

\noindent We have proposed a novel cooling, trapping, and slowing scheme for neutral particles, which we have termed the SWAP MOT. We have outlined an experimental protocol and developed a simple and efficient semiclassical model and simulation that demonstrates its ability to generate phase space compression. We have also specified a protocol for determining appropriate experimental parameters for application to an arbitrary transition. Our results demonstrate the SWAP MOT's ability to generate large coherent and dissipative forces with fewer scattered photons than traditional MOTs. We provided a discussion of the resulting equilibrium 1D temperature and cloud size scaling properties with Rabi and recoil frequencies and compared the results to those found from our theoretical model of the traditional MOT.

Aside from eventual experimental implementation, there are several topics that could be studied in future work. We have not explicitly simulated 3D dynamics, but we conjecture that applying the SWAP MOT protocol to each dimension successively in time is a viable option~\cite{snigirev}. Also, the incorporation of shortcuts to adiabaticity could speed up the protocol with a low energetic cost \cite{shortcut}. Moreover, there are a number of additions to the simulation that would be required to more accurately capture the dynamics in the quantum regime (aside from developing a computationally efficient algorithm to simulate the fully quantum model developed in this work), such as spontaneous photon reabsorption, particle collisions, and system loss to states that are not resonant with the cooling laser frequencies.

Furthermore, we identify that the stimulated emission process can be realized in our model via any operation that reverses the sign of the magnetic dipole potential energy
\begin{equation}
    \hat{H}_B(\hat{x}) = - \bs{\hat{\mu}} \cdot \bs{\hat{B}}(\hat{x})
\end{equation}
at the middle of the sweep, where 
\begin{equation}
        \bs{\hat{\mu}} = - \frac{g_J \mu_\text{B}}{\hbar} \bs{\hat{J}}
\end{equation}
is the magnetic dipole moment of the particle and $\bs{\hat{J}}$ is the total electronic angular momentum. We incorporated the sign flip by magnetic field switching, but this is technically difficult in an experimental setting and it is fundamentally impossible to perfectly, diabatically switch the magnetic field direction. Other potentially more experimentally viable options are applying $\pi$-pulses at the middle of each sweep in order to transfer the population between the excited Zeeman sublevels, thereby flipping the sign of $\bs{\hat{J}}$, or optically pumping into a nearby hyperfine manifold with a spin $g$-factor of opposite sign.

We   would   like   to   thank   Jun   Ye,   Shiqian   Ding, Yewei Wu, Ian Finneran, Matthew Norcia, and Athreya Shankar  for  many  useful  discussions. This  work  was supported by NSF grant number PHY 1806827 and the DARPA Extreme Sensing Program.

\bibliographystyle{apsrev4-1}
\bibliography{SCT.bib}

\appendix

\section{SWAP MOT Effective Potential} \label{dimple}

\noindent We have shown in Section~\ref{equilibrium} that the equilibrium position distribution in a SWAP MOT $X(x)$ is well-described by the sum of two Gaussian functions with different widths:
\begin{align}
\label{eq:positionDist}
    X(x) \approx
    &\frac{w}{\sqrt{2 \pi} \sigma_1}
    \exp \left(-
    \frac{x^2}{2\sigma_1^2}
    \right)\notag \\
    +
    &\frac{(1-w)}{\sqrt{2 \pi} \sigma_2}
    \exp \left(-
    \frac{x^2}{2\sigma_2^2}
    \right),
\end{align}
where $0\leq w \leq 1$ describes the relative weight of the two Gaussians, and $\sigma_1$ and $\sigma_2$ are the widths of the two distributions. Note that we have normalized the distribution for simplicity. In this Appendix, we derive an effective potential $V_\text{eff}(x)$ that characterizes the resulting position distribution due to the effects of the SWAP MOT.

In a state of thermodynamic equilibrium, the  classical phase space distribution $\rho$ of a system is given by \cite{statMech}
\begin{equation}
    \rho(x,p) =
    Z^{-1}e^{-H(x,p)/k_BT},
\end{equation}

\noindent where $Z$ is the partition function and $H$ is the system Hamiltonian. If we consider a system with a Hamiltonian of the form
\begin{equation}
    H(x,p) = \frac{p^2}{2m} + V_\text{eff}(x),
\end{equation}

\noindent where $V_\text{eff}(x)$ is an effective potential, then the position distribution $X(x)$ satisfies
\begin{equation}
    X(x) \equiv \int \rho(x,p) \, dp
    \propto  e^{-V_\text{eff}(x)/k_BT}.
\end{equation}

\noindent Equivalently, 
\begin{equation}
    V_\text{eff}(x) \propto - \log X(x).
\end{equation}

In the case of the SWAP MOT, we find that
\begin{align}
\label{eq:vEFf}
    V_\text{eff}(x) \propto -
    \log
    &\biggl{[}
    \frac{w}{\sqrt{2 \pi} \sigma_1}
    \exp \left(-
    \frac{x^2}{2\sigma_1^2}
    \right)\notag \\
    &+
    \frac{(1-w)}{\sqrt{2 \pi} \sigma_2}
    \exp \left(-
    \frac{x^2}{2\sigma_2^2}
    \right)
	\biggr{]}
\end{align}
by using Eq~(\ref{eq:positionDist}). We plot this effective potential against the simulated position space distribution arising in the SWAP MOT in Fig.~\ref{fig:eq}(f).

\end{document}